# Obtaining and Application of the Reduced Bloch Equations

**M.D. Zviadadze, N.D.Chachava, A.G. Kvirikadze, A.K. Pokleba**
E. Andronikashvili Institute of Physics, 0177. Tamarashvili 6. Tbilisi Georgia

e-mail:   m.zviadadze@mail.ru,    m.zviadadze@aiphysics.ge

**Abstract**

Well-known Bloch equations describe the spin systems (electronic and nuclear) for any scale of time, from transient processes to steady states. Usually in solids $T_2 \ll T_1$. The question arises: what are the approximations that should be made in order the "roughen" Bloch equations to describe the processes at the time $T_2 < t < T_1$? The answer to this question is given in this article. As an example, the saturation of magnetic resonance is considered under the conditions of harmonic modulation of a constant magnetic field.

1. Phenomenological theory of magnetic resonance is based on the well-known Bloch equations [1]:

$$\frac{d\vec{m}(t)}{dt} = \gamma \vec{m}(t) \times \vec{H}(t) - \frac{m_x(t)\vec{i} + m_y(t)\vec{j}}{T_2} - \frac{m_z(t) - m_0}{T_1}\vec{k}, \qquad (1)$$

where $\vec{m}(t)$ is the vector of magnetization, $\gamma$ is the gyromagnetic ratio, $T_1$ and $T_2$ are the times of the spin-spin (transverse) and the spin-lattice (longitudinal) relaxations, $m_0$ is the equilibrium value of magnetization, $\vec{i}, \vec{j}, \vec{k}$ are the orts of a laboratory coordinate system (LCS), $\vec{H}(t)$ is the alternating magnetic field. Further, as a specific example, we shall consider the magnetic field:

$$H_x(t) = 2H_1 \cos\Omega t,\; H_y(t) = 0,\; H_z(t) = H_0 + H_m \cos\omega t, \qquad (2)$$

where $2H_1, \Omega$ are the amplitude and the frequency of linearly polarized transverse field, $H_0$ is the constant magnetic field, $H_m, \omega$ are the amplitude and the frequency of modulation. In this case, equation (1) describes the phenomenon of magnetic resonance taking into account the longitudinal modulation.

In terms of spherical components of the magnetization $m^\pm = m_x \pm im_y$, $m_z$ and of the field $H^\pm = H_x \pm iH_y$, $H_z$, set of equations (1) is transformed into the following form:

$$\begin{cases} \dfrac{dm^+(t)}{dt} + i(\omega_0 + \Delta_m \cos \omega t)m^+(t) + \dfrac{m^+(t)}{T_2} = 2i\omega_1 \cos \Omega t\, m_z(t) \\ \dfrac{dm_z(t)}{dt} = -i\omega_1 \cos \Omega t\, \mathrm{Im}(m^+(t)) - \dfrac{m_z(t) - m_0}{T_1}, \qquad m^-(t) = \{m^+(t)\}^* \end{cases} \tag{3}$$

where $\omega_0 = \gamma H_0$ is the Zeeman frequency, $\omega_1 = \gamma H_1$, $\Delta_m = \gamma H_m$ and the sign $\{...\}^*$ denotes the complex conjugation.

Bloch equations describe the spin systems (electronic and nuclear) in arbitrary magnetic fields and at all time scales, from transient processes to steady states. In solids $T_2 \ll T_1$, so, the question arises: how to "roughen" and thus how to simplify these equations in order they to describe the processes at the scale of time $T_2 < t < T_1$ only. The answer to this question can be obtained using the method proposed in [2].

**2.** Performing the transformation

$$\tilde{m}^\pm(t) = m^\pm(t) \exp(\pm \Omega t)$$

let us pass to the rotating coordinate system (RCS), which rotates around the $z$ axis with frequency $\Omega$. Dropping, as usual, the non-resonance terms containing the frequency $2\Omega$, we obtain a set of Bloch equations in the RCS:

$$\begin{cases} \dfrac{d\tilde{m}^+(t)}{dt} + \left(i\Delta(t) + \dfrac{1}{T_2}\right)\tilde{m}^+(t) = i\omega_1 m_z(t), \\ \dfrac{dm_z(t)}{dt} = -\omega_1 \mathrm{Im}(\tilde{m}^+(t)) - \dfrac{m_z(t) - m_0}{T_1}, \qquad \tilde{m}^-(t) = \{\tilde{m}^+(t)\}^*; \end{cases} \tag{4}$$

where $\Delta(t) = \Delta_0 + \Delta_m \cos \omega t$, $\Delta_0 = \omega_0 - \Omega$ is the resonance detuning.

Let's introduce a new unknown function:

$$m'^+(t) = \tilde{m}^+(t) \exp(ia \sin \omega t),\ a = \Delta_m / \omega. \tag{5}$$

Transformation (5) makes a transition to the $R$-system (in our terminology) and allows to simplify significantly the study of the role of modulation. So, set of equations (4) takes the form:

$$\begin{cases} \dfrac{dm'^+(t)}{dt} + \left(i\Delta_0 + \dfrac{1}{T_2}\right)m'^+(t) = i\omega_1(t)m_z(t) \\ \dfrac{dm_z(t)}{dt} = -\mathrm{Im}(\omega_1^*(t)m'^+(t)) - \dfrac{m_z(t) - m_0}{T_1}, \qquad m'^-(t) = \{m'^+(t)\}^* \end{cases}, \tag{6}$$

where the following notation is introduced:

$$\omega_1(t) = \omega_1 \exp(ia \sin \omega t). \tag{7}$$

Sets of equations (6) and (4) are fully equivalent, but they differ in interpretation of modulation if using the well-known expansion

$$exp(\pm ia\sin\omega t) = \sum_{i=-\infty}^{+\infty} J_k(a)exp(\pm ik\omega t), \quad (8)$$

where $J_k(a)$ are Bessel functions of the whole order. In (4), there is one constant transverse magnetic field with amplitude $\omega_1$ (in frequency units) and the instant detuning of the resonance $\Delta(t)$ varies with the frequency of modulation $\omega$; whereas in (6) detuning of the resonance is constant, but there are many transverse fields with frequencies $k\omega (k \in Z)$. Obviously, $|\omega_1(t)| = \omega_1$, therefore, the longitudinal modulation $\Delta_m, \omega$ does not affect the magnitude of interaction between the spin system (SS) and the transverse field, so if $\omega_1 \ll T_2^{-1}$ (an intermediate saturation of magnetic resonance, the so-called Provotorov's case [3]), the interaction with transverse field remains small at any $\Delta_m, \omega$.

3. Let's establish the relationship between the transverse $m'^{\pm}(t)$ and longitudinal $m_z(t)$ components of magnetization. According to [2] and considering the right-hand part in the first equation of set (6) as an inhomogeneous term, it is easy to obtain the general solution of this equation:

$$m'^+(t) = m_0'^+(t) + i\int_0^t dt_1 \, exp\left[-\left(i\Delta_0 + \frac{1}{T_2}\right)t_1\right]\omega_1(t-t_1)m_z(t-t_1), \quad (9)$$

where

$$m'^+(t) = m_0^+(0)\cdot exp\left[-\left(i\Delta_0 + \frac{1}{T_2}\right)t\right] \quad (10)$$

is the general solution of homogeneous equation at the initial condition $m'^+(t)\big|_{t=0} = m'^+(0) = m^+(0)$. Thus, $m'^+(t)$ functionally depends on all previous values of longitudinal magnetization $m_z(t-t_1)(0 \le t \le t_1)$, i.e. Bloch equations (3) and (4) with $\omega_1 \ne 0$ initially contain implicitly the memory.

According to (5),

$$\tilde{m}^+(t) = \tilde{m}_0^+(t) + i\omega_1\int_0^t dt_1 \, f(t_1)m_z(t-t_1)exp[-i\Delta_0 t_1 + ia(\sin\omega(t-t_1) - \sin\omega t)], \quad (11)$$

where

$$\tilde{m}_0^+(t) = m^+(0)exp\left[-\left(i\Delta_0 + \frac{1}{T_2}\right)t - ia\sin\omega t\right] \quad (12)$$

and a spin correlation function $f(t_1) = exp(-|t_1|/T_2)$ is introduced. Fourier-image of this function is the Lorentz form of magnetic resonance line of $2/T_2$ width at the half-height, as it should be when using the Bloch equations. Substituting solution (11) in the second equation of set (4) and using the relation $Im(iz) = Re(z)$, we obtain the equation:

$$\frac{dm_z(t)}{dt} = \omega_1 Im(\tilde{m}_0^+(t)) - \int_0^t dt_1 f(t_1) m_z(t-t_1) Re[\exp(-i\Delta_0 t_1) \omega_1^*(t) \omega_1(t-t_1)] - \frac{m_z(t) - m_0}{T_1}. \quad (13)$$

By the method given in [2], one can transform the integro-differential equation (13) into the Volterra integral equation of the second kind with complex non-difference kernel. This equation, in principle, can be solved by the method of successive approximation, but even in the simplest cases, we obtain very cumbersome results, practically unsuitable for the analysis. Therefore, confine ourselves to the approximated solution of equation (13).

**4**. Equations (11), (13), taking into account (12), are fully equivalent to the set of Bloch equations in the RSC. "Roughen" Bloch equations describing SS at $t \gg T_2$ time, can be obtained from the expression (11) and from the following considerations:

- When $t \gg T_2$, the solution of homogeneous equation (12) vanishes - the transient effects disappear. In addition, in expression (11) under the sign of the integral there is a "cut-off" factor $\exp(-t_1/T_2)$, so the upper limit of integration can be replaced by $\infty$.

- In solids, $T_2 \ll T_1$. In strong variable magnetic fields $\omega_1 \gg T_2^{-1}$ ($\varepsilon = (1/T_2 - 1/T_1)/\omega \ll 1$ in the notations of Torrey [4]) the efficient relaxation times $\tilde{T}_1$ and $\tilde{T}_2$ (in our notation) are given by the expressions:

$$\tilde{T}_1^{-1} = T_1^{-1} \cos^2\theta + T_2^{-1} \sin^2\theta, \quad \tilde{T}_2^{-1} = (2T_2)^{-1}(1 + \cos^2\theta) + (2T_1)^{-1} \sin^2\theta,$$

where $\sin\theta = \omega_1/\omega_e$, $\omega_e = \sqrt{\Delta_0^2 + \omega_1^2} = H_e/\gamma$ and $\theta$ is the angle between $z$ axis and the efficient magnetic field $\vec{H}_e = \omega_1 \vec{i} + \Delta_0 \vec{k}$, acting on SS in RCS. When $\omega_1 \ll T_2^{-1}$, i.e., $\theta \ll 1$, we obtain $\tilde{T}_1 \approx T_1$ and $\tilde{T}_2 \approx T_2$. If we introduce the dimensionless time $\tau = t/T_2$ in equations (11) and (13), it is easy to ascertain that $\frac{d\tilde{m}^+(\tau)}{d\tau} \sim \omega_1 \tau$ and $\frac{dm_z(\tau)}{d\tau} \sim \omega_1^2 T_2^2$. Thus, for not very strong variable fields ($\omega_1 \ll T_2^{-1}$) $\frac{dm_z(\tau)}{d\tau} \ll \frac{d\tilde{m}^+(\tau)}{d\tau}$, so, the transverse components of magnetization $m^\pm(t)$ remained "fast", and the longitudinal component $m_z(t)$ - "slow", therefore, in (11) we can make the replacement $m_z(t-\tau) \to m_z(t)$. In the case of arbitrary values of $H_1$, the relation between $\tilde{T}_1$ and $\tilde{T}_2$ can be of any type.

Following the made comments, the equations (11) and (13) become:

$$\tilde{m}^+(t) = K^+(t) m_z(t),$$

$$K^+(t) = i\omega_1 \int_0^\infty dt_1 f(t_1) \exp[-i\Delta_0 t_1 + ia(\sin\omega(t-t_1) - \sin\omega t)], \quad (14)$$

$$\frac{dm_z(t)}{dt} = -\omega_1^2 \int_0^\infty dt_1 f(t_1) Re\{\exp[i\Delta_0 t_1 + ia(\sin\omega(t-t_1) - \sin\omega t)]\} m_z(t) - \frac{m_z(t) - m_0}{T_1}. \quad (15)$$

**5.** Let us consider small modulation frequencies $\omega \ll T_2^{-1}$. In this case, in formula (14) we can use the expansion

$$\sin \omega(t - t_1) \cong \sin \omega t - \omega t_1 \cos \omega t$$

and the value $K^+(t)$ takes the form

$$K^+(t) = i\omega_1 \int_0^\infty dt_1 f(t_1) \operatorname{Re} \exp[-i\Delta(t)t_1] = i\pi\omega_1 f(\Delta(t)), \quad (16)$$

where $f(\Delta(t)) = \{\pi T_2 [\Delta^2(t) + T_2^{-2}]\}^{-1}$ is the shape of a line with instant detuning, and $T_2 = \pi f(0)$.

To analyze the obtained results, it is convenient to introduce the dimensionless time $\tau = \omega t$.

As a result, for $m_z(\tau)$ we obtain the following equation:

$$\frac{dm_z(\tau)}{d\tau} = -\frac{2W(\Delta(\tau))}{\omega} m_z(\tau) - \frac{m_z(\tau) - m_0}{T_1}, \quad (17)$$

where $2W(\Delta(\tau)) = \pi\omega_1^2 f(\Delta(\tau)) = \omega_1^2 T_2 \frac{f(\Delta(\tau))}{f(0)}$.

It is well-known that SS of solid state at $t \gg T_2$ in the approximation of high temperatures and under the conditions of intermediate resonance $(\omega_1 \ll T_2^{-1})$ is described by Provotorov's equations [3] for inverse temperatures of $\beta_z(t)$ Zeeman sub-system and for dipole reservoir (DR) $\beta_d(t)$. If DR quickly comes into equilibrium with the lattice $(\beta_d = \beta_L)$, or if spin concentrations are too small to take DR into account (formally, this means that $\beta_d = 0$), Provotorov's equations reduce to one equation (17) with the only difference that the correlation function $f(t)$, generally speaking, is not of Lorentz form, and the value $W(\Delta(\tau))$ plays the role of the probability of spin flip by the transverse magnetic field.

The general solution of Eq. (17) at the initial condition $m_z(\tau)|_{\tau=0} = m_z(0)$ has the following form:

$$m_z(\tau) = m_{z0}(\tau) + \frac{m_0}{\omega T_1} \int_0^\tau d\tau_1 \exp\left(-\int_{\tau_1}^\tau \frac{d\tau_2}{\omega T_R(\tau_2)}\right), \quad (18)$$

where the term

$$m_{z0}(\tau) = m_z(0) \exp\left(-\int_0^\tau \frac{d\tau_1}{\omega T_R(\tau_1)}\right) \quad (19)$$

is the general solution of homogeneous equation $m_0 = 0$, and in (17) the following notation is introduced.

$$T_R^{-1}(\tau) = T_1^{-1} + 2W(\Delta(\tau)) = T_1^{-1} + \pi\omega_1^2 f(\Delta(\tau)). \quad (20)$$

In accordance with the theorem of Floquet, expression (19) can be given the following

$$m_{z0}(\tau) = \exp(-K\tau) \cdot F(\tau), \tag{21}$$

where $K$ is the constant number, $F(\tau)$ is the periodic function ((21) is convenient for making the further calculations and for the analysis of the obtained results). This can be achieved using the following transformations. The value $f(\Delta(\tau))$ is the periodic function of dimensionless time with the period of $\tau_n = 2\pi$ (true modulation period is equal to $T = 2\pi/\omega$), if $\Delta_0 \neq 0$, and with the period of $\tau_m = \pi$ at $\Delta_0 = 0$, therefore, it can be expanded into Fourier series

$$f(\Delta(\tau)) = \sum_{n=-\infty}^{\infty} f_n \exp(-iK\tau), \quad f_n = \frac{1}{\tau_m} \int_0^{\tau_m} d\tau f(\Delta(\tau)) \exp(in\tau).$$

Fourier coefficient

$$f_{n=0} = \frac{1}{\tau_m} \int_0^{\tau_m} f(\Delta(\tau))d\tau = \bar{f} \tag{22}$$

is the average value of $f(\Delta(\tau))$ function over the dimensionless modulation period $\tau_m$. The difference

$$f(\Delta(\tau)) - \bar{f} = \sum_{n \neq 0} f_n \exp(-in\tau) \tag{23}$$

does not contain the constant term anymore, therefore, we make the transformation of (20) in the following way:

$$T_R^{-1}(\tau) = T_e^{-1} + \tau_s^{-1}(0)[f(\Delta(\tau)) - \bar{f}]/f(0), \tag{24}$$

where the value

$$T_e^{-1} = T_1^{-1} + \pi\omega_1^2 \bar{f} = T_1^{-1} + \tau_s^{-1}(0)\bar{f}/f(0) \tag{25}$$

is the efficient decay time of longitudinal magnetization, and where the notation $\tau_s^{-1}(0) = \omega_1^2 T_2 = 2W(0)$ is introduced. The parameter $\tau_s^{-1}(0)$ is the rate with which the saturation is reached in the center of resonance curve [5]. Further on, it will be considered that the relation $\tau_s(0)/T_1 = (\omega_1^2 T_1 T_2)^{-1} \equiv s^{-1}(0) \ll 1$, where $s(0) = \omega_1^2 T_1 T_2$ – is the parameter of saturation in the center of resonance curve. Thus, the decay of longitudinal magnetization is caused by two reasons: by saturation (fast process) and by spin-lattice relaxation (slow process).

Taking into account Eq. (25), we find

$$\int_0^{\tau} \frac{d\tau_1}{\omega T_R(\tau_1)} = \tau/(\omega T_e) + \Phi(\tau)/(\omega \tau_s(0)), \tag{26}$$

where the odd periodic function is introduced

$$\Phi(\tau) = \int_0^\tau d\tau_1 \left[ f(\Delta(\tau_1)) - \bar{f} \right] / f(0) \tag{27}$$

with $\tau_m$ period. As a result, the function $m_{z0}(\tau)$ takes the form:

$$m_{z0}(\tau) = m_z(0) \exp\{-\tau/(\omega T_e) + \Phi(\tau)/(\omega \tau_s(0))\}, \tag{28}$$

which coincides with (21) at $K = (\omega T_e)^{-1}$ and $F(\tau) = \exp\{\Delta(\tau)/(\omega \tau_s(0))\}$[1]. As

$$\exp\left\{-\int_{\tau_1}^\tau \frac{d\tau_2}{\omega T_R(\tau_2)}\right\} = \exp\left\{-\frac{\tau - \tau_1}{\omega T_e} - \frac{\Phi(\tau) - \Phi(\tau_1)}{\omega \tau_s(0)}\right\},$$

the general solution of non-homogeneous equation (17) after changing the variable $\tau - \tau_1 \to \tau_1$, is given by expression:

$$m_z(\tau) = m_{z0}(\tau) + \frac{m_0}{\omega T_1} \int_0^\tau d\tau_1 \exp\left\{-\tau_1/(\omega T_e) - \frac{\Phi(\tau) - \Phi(\tau - \tau_1)}{\omega \tau_s(0)}\right\}. \tag{29}$$

The steady state solution $m_{z,st}(\tau) = m_z(\tau)\big|_{t \gg T_e}$ is obtained from (29) using the limiting transition

$$m_{z,st}(\tau) = \lim_{\tau \to \infty} m_z(\tau) = \frac{m_0}{\omega T_1} \int_0^\infty d\tau_1 \exp\left\{-\tau_1/(\omega T_e) - \frac{\Phi(\tau) - \Phi(\tau - \tau_1)}{\omega \tau_s(0)}\right\}. \tag{30}$$

Obviously, $m_{z,st}(\tau + \tau_m) = m_{z,st}(\tau)$. Formula (30) allows the further simplification. Let us present the integral in (30) as:

$$\int_0^\infty d\tau_1 (\ldots) = \sum_{n=0}^\infty \int_{n\tau_m}^{(n+1)\tau_m} d\tau_1 (\ldots).$$

Then

$$m_{z,st}(\tau) = \frac{m_0}{\omega T_1} \sum_{n=0}^\infty \int_{n\tau_n}^{(n+1)\tau_m} d\tau_1 \exp\left\{-\frac{\tau_1}{\omega T_e} - \frac{\Phi(\tau) - \Phi(\tau - \tau_1)}{\omega \tau_s(0)}\right\}.$$

We introduce a new variable $\tau_2 = \tau_1 + n\tau_m$ and, as a result, obtain the expression

$$m_{z,st}(\tau) = \frac{m_0}{\omega T_1} \sum_{n=0}^\infty \int_0^{\tau_m} d\tau_1 \exp\left\{-\frac{\tau_1 + n\tau_m}{\omega T_e} - \frac{\Phi(\tau) - \Phi(\tau - \tau_1)}{\omega \tau_s(0)}\right\}$$

---

[1] Expression (28) can be obtained if we seek the solution of homogeneous equations in the form of (21) and determine $K$ from the condition of periodicity of $F(\tau)$ function.

.

After summing the infinite decreasing geometric progression with the denominator $q = exp(-\tau_m/(\omega T_e))$, we finally find:

$$m_{z,st}(\tau) = \frac{m_0}{\omega T_1}[1 - exp(-\tau_m/(\omega T_e))]^{-1} \int_0^{\tau_m} d\tau_1 \, exp\left\{-\frac{\tau_1}{\omega T_e} - \frac{\Phi(\tau) - \Phi(\tau-\tau_1)}{\omega \tau_s(0)}\right\}. \quad (31)$$

Let us analyze the final solution of (31) under the conditions of deep modulation [2)]

$$\Delta_m \gg T_2^{-1}, \quad T_2\left|\frac{d\Delta(t)}{dt}\right| \ll T_2^{-1} \Rightarrow \Delta_m \ll (\omega T_2^2)^{-1} \quad (32)$$

During the dimensionless period of modulation $\tau_m = 2\pi$ (for definiteness assume that $\Delta_0 \neq 0$), the condition of exact resonance $\Delta(\tau) = 0$ is fulfilled twice at the moments of time $\tau_1$ and $\tau_2 = 2\pi - \tau_1$ which are the roots of the equation $\Delta_0 + \Delta_m \cos\tau = 0$ (at $\Delta_0 = 0$ $\tau_1 = \pi/2$, $\tau_2 = 3\pi/2$). At the deep modulation for period $2\pi$ the saturation of resonance is reached twice in the vicinity of $\tau_1$ and $\tau_2$ moments, and the duration of saturation process $\tau_p$ is determined from the condition $|\Delta(\tau)| \leq T_2^{-1}$, and its order of magnitude appears to be equal to $\tau_p \sim 2\pi/(\Delta_m T_2) \ll 2\pi$. Obviously, $\tau_p$ is of the order of time of passing the resonance line and it should be greater, than the time of establishment the saturation $\tau_s(0)$, i.e. $\tau_p > \tau_s(0)$. If we want to consider SS as isolated from the lattice during the process of saturation, it is necessary to fulfill the condition $\tau_s(0) \ll \omega T_1$, i.e. $(\omega T_1)^{-1} \ll \Delta_m T_2$. Thus, for $2\pi$ period of time the duration of saturation is $\sim 2\tau_p$, and the duration of $S-L$ relaxation $\sim 2\pi - 2\tau_p \approx 2\pi$. We assume that $\omega \sim T_1^{-1}$. In this case, in the intervals between the saturations, the equilibrium state is almost reached in SS.

Qualitatively, the character of changing $m_{z,st}(\tau)$ under the conditions of deep modulation is as follows: during $\tau_m = 2\pi$, in the vicinity of $\tau_1$ and $\tau_2$ moments, SS is subjected to two narrow saturated pulses of $\tau_p \ll 2\pi$ duration, and besides, during the action of $\tau_1$-pulse, the resonance line is passed from left to the right $(\Omega < \omega_0 \to \Omega > \omega_0)$, and during $\tau_2$-pulse, on the contrary, the line is passed from the right to the left $(\Omega > \omega_0 \to \Omega < \omega_0)$. The distance $\tau_2 - \tau_1$ between pulses can be changed by resonance detuning $\Delta_0$. In the saturation periods, $m_{z,st}(\tau)$ changes quickly (with the rate of $\sim W$) from the equilibrium value $m_0$ to the saturated one $\sim \frac{m_0}{s} \approx 0$, while in the of $S-L$ relaxation periods it changes (with the rate of $\sim T_1^{-1} \ll W$) and the equilibrium value $m_0$ has enough time to be established or almost established. Thus, at the deep modulation, the longitudinal Let us go from the qualitative analysis to the quantitative calculation. In $S-L$ relaxation region, in (31) we can insert $\Phi(\tau) \approx 0$, that gives

$$m_{z,st} = m_0 T_e/T_1 = m_0(1 + s\bar{f}/f(0))^{-1}. \quad (33)$$

---

[2)] The second inequality in (32) means the infinitesimal change of resonance detuning $\Delta(t)$ for $T_2$ time, as compared to the line width which is $\sim T_2^{-1}$. Otherwise, the saturation of resonance cannot be reached magnetization undergoes relaxation oscillations [5].

As, always $\bar{f} < f(0)$, from (33) it follows that modulation makes the saturation weaker. At $\Delta_0 = 0$ the integral (22) is taken exactly:

$$\bar{f} = \frac{1}{\pi}\int_0^\pi f(\Delta_m \cos\tau)d\tau = f(0)/\sqrt{1+\Delta_m^2 T_2^2}\ .$$

At the deep modulation

$$\bar{f}/f(0) \approx (\Delta_m^2 T_2^2)^{-1} \ll 1. \tag{34}$$

In the general case $\Delta_0 \neq 0$, it is impossible to calculate exactly the integral (22), however the estimation (34) remains valid.

In saturation region we can neglect $S - L$ relaxation and (31) gives

$$m_{z,st}(\tau) = \frac{m_0}{\omega T_1}\left[1-\exp\left(-\frac{\tau_m}{\omega T_e}\right)\right]^{-1}\int_0^{\tau_m} d\tau_1\, \exp\left\{-\frac{\Phi(\tau)-\Phi(\tau-\tau_1)}{\omega\tau_s(0)}\right\}. \tag{35}$$

At the deep modulation, the saturation region $\sim (\Delta_m T_2)^{-1} \ll \tau_m$, therefore in (35) we can make expansion in terms of $\tau_1$:

$$\Phi(\tau-\tau_1) \cong \Phi(\tau) - \frac{f(\Delta(\tau))-\bar{f}}{f(0)}\tau_1.$$

As a result, we obtain

$$m_{z,st}(\tau) \cong \frac{m_0}{\omega T_1}\left[1-\exp\left(-\frac{\tau_m}{\omega T_e}\right)\right]^{-1}\int_0^{\tau_m} d\tau_1\, \exp\left\{-\frac{f(\Delta(\tau))-\bar{f}}{f(0)\tau_s(0)}\tau_1\right\}. \tag{36}$$

The different aspects of the action of longitudinal modulation on SS described by Bloch equations, were discussed in [2] theoretically and experimentally. In particular, the authors managed to observe the abrupt changes $m_z(t)$, similar to the relaxation oscillations.

**6**. Let's consider linear approximation with respect to amplitude of modulation $H_m$ (in this case no limitations are imposed on frequency $\omega$). Correspondingly, in formula (15) the exponent under the integral can be changed by the following expression

$$\exp[ia(\sin\omega(t-t_1)-\sin\omega t)] \cong 1 + ia(\sin\omega(t-t_1)-\sin\omega t).$$

Representing $\exp(i\Delta_0 t_1)$ according to the Euler formula and taking the real part of the exponent, we get

$$\frac{dm_z(t)}{dt} = -\omega_1^2\int_0^\infty dt_1 f(t_1)[\cos\Delta_0 t_1 - a\sin\Delta_0 t_1(\sin\omega(t-t_1)-\sin\omega t)]m_z(t) - \frac{m_z(t)-m_0}{T_1}. \tag{37}$$

Introducing the notations

$$S(x) \equiv \int_0^\infty d\tau f(\tau)\sin x\tau, \ C(x) \equiv \int_0^\infty d\tau f(\tau)\cos x\tau \tag{38}$$

equation (37) takes the following form:

$$\frac{dm_z(t)}{dt} = -\omega_1^2 m_z(t)C(\Delta_0) - \frac{m_z(t)-m_0}{T_1} - \frac{1}{2}a\omega_1^2 m_z(t) \cdot$$
$$\cdot\{\sin\omega t[2S(\Delta_0) - S(\Delta_0+\omega) - S(\Delta_0-\omega)] - \cos\omega t[C(\Delta_0-\omega) - C(\Delta_0+\omega)]\}. \tag{39}$$

Let's seek the steady solution of equation (39) ($t \gg T_1$), as follows:

$$m_z(t) = A\sin\omega t + B\cos\omega t + C. \tag{40}$$

Substituting (40) in (39) and considering that the free terms and the coefficients in front of the sine and cosine must vanish separately, we get a set of equations:

$$\begin{cases} m_0 - C(1+\omega_1^2 C(\Delta_0)T_1) = 0, \\ A\left(\frac{1}{T_1} + \omega_1^2 C(\Delta_0)\right) - B\omega + \frac{C a\omega_1^2}{2}[2S(\Delta_0) - S(\Delta_0+\omega) - S(\Delta_0-\omega)] = 0, \\ A\omega + B\left(\frac{1}{T_1} + \omega_1^2 C(\Delta_0)\right) + \frac{C a\omega_1^2}{2}[C(\Delta_0-\omega) - C(\Delta_0+\omega)] = 0. \end{cases} \tag{41}$$

Solution of this set of equations is:

$$A = \frac{\beta_1 \alpha + \beta_2 \omega}{\alpha^2 + \omega^2}; \ B = \frac{\beta_2 \alpha - \beta_1 \omega}{\alpha^2 + \omega^2}; \ C = \frac{m_0}{1+\omega_1^2 C(\Delta_0)T_1}, \tag{42}$$

) where the following notations are used:

$$\begin{cases} \alpha \equiv \frac{1}{T_1} + \omega_1^2 C(\Delta_0), \ \beta_1 \equiv -\frac{C a\omega_1^2}{2}[2S(\Delta_0) - S(\Delta_0+\omega) - S(\Delta_0-\omega)] \\ \beta_2 \equiv -\frac{C a\omega_1^2}{2}[C(\Delta_0-\omega) - C(\Delta_0+\omega)]. \end{cases} \tag{43}$$

Let us introduce the difference in magnetizations

$$\Delta m_z(t) = m_z(t) - m_{z0} = Re[\chi \exp(i\omega t)], \tag{44}$$

where

$$m_{z0} = m_z(t)|_{t=0} = C = \frac{m_0}{1+\omega_1^2 C(\Delta_0)T_1} \tag{45}$$

is the steady magnetization of SS in the absence of modulation, and $\chi = \chi' - i\chi''$ is the complex susceptibility. Obviously:

$$\Delta m_z(t) = H_m(\chi' \cos \omega t + \chi'' \sin \omega t). \tag{46}$$

The comparison of (40) and (46) gives:

$$A = H_m \chi'', \quad B = H_m \chi',$$

and leads to the following expressions for longitudinal magnetization:

$$\chi' = \frac{\gamma \omega_1^2}{2\omega} \frac{m_{z0}}{\omega^2 + \left(\frac{1}{T_1} + \omega_1^2 C(\Delta_0)\right)^2} \cdot$$

$$\cdot \left\{ \left(\frac{1}{T_1} + \omega_1^2 C(\Delta_0)\right) [C(\Delta_0 - \omega) - C(\Delta_0 + \omega)] - \omega[2S(\Delta_0) - S(\Delta_0 + \omega) - S(\Delta_0 - \omega)] \right\};$$

$$\chi'' = \frac{\gamma \omega_1^2}{2\omega} \frac{m_{z0}}{\omega^2 + \left(\frac{1}{T_1} + \omega_1^2 C(\Delta_0)\right)^2} \cdot \tag{47}$$

$$\cdot \left\{ \omega[C(\Delta_0 - \omega) - C(\Delta_0 + \omega)] + \left(\frac{1}{T_1} + \omega_1^2 C(\Delta_0)\right) [2S(\Delta_0) - S(\Delta_0 + \omega) - S(\Delta_0 - \omega)] \right\}$$

We could solve the Bloch equations (1) in linear approximation with respect to $H_m$. In doing so, the complex expressions are obtained, being inconvenient to work with. Then, if we consider the time $t \gg T_1$ and limit ourselves with the lowest approximations with respect to the parameter $\omega_1 T_2 \ll 1$, we shall obtain the expressions similar to (47), which were found from the roughen Bloch equations.

It is easy to see that the relation $\chi'(-\omega) = \chi'(\omega)$, $\chi''(-\omega) = -\chi''(\omega)$, is in accord with the general theory of generalized susceptibility [5]. It can be easily shown that for Lorentz correlation function $f(t) = \exp(-|t|/T_2)$, the expressions (38) have the following form:

$$C(x) = \pi f(x), \quad S(x) = \pi x T_2 f(x), \quad f(x) = \frac{T_2}{\pi} \frac{1}{1 + x^2 T_2^2} \tag{48}$$

here $\chi'$, $\chi''$ can be written as:

$$\chi'(\omega) = \frac{m_0 \gamma T_1}{2\omega} \frac{1}{\omega^2 T_1^2 + (1 + s(\Delta_0))^2} \cdot$$

$$\left\{ [W(\Delta_0 + \omega) - W(\Delta_0 - \omega)] + \frac{\omega T_1 T_2}{1 + s(\Delta_0)} [2\Delta_0 W(\Delta_0) - (\Delta_0 + \omega)W(\Delta_0 + \omega) - (\Delta_0 - \omega)W(\Delta_0 - \omega)] \right\};$$

$$\chi'(\omega) = \frac{m_0 \gamma T_1}{2\omega} \frac{1}{\omega^2 T_1^2 + (1 + s(\Delta_0))^2} \cdot \tag{49}$$

$$\cdot \left\{ \frac{\omega}{1 + s(\Delta_0)} [W(\Delta_0 + \omega) - W(\Delta_0 - \omega)] - \frac{T_2}{T_1} [2\Delta_0 W(\Delta_0) - (\Delta_0 + \omega)W(\Delta_0 + \omega) - (\Delta_0 - \omega)W(\Delta_0 - \omega)] \right\};$$

$s(\Delta_0) = \pi \omega_1^2 T_1 f(\Delta_0)$ is the parameter of saturation with detuning $\Delta_0$, $s(0) = \omega_1^2 T_1 T_2$ is the parameter of saturation in the center of line, and $W(x) = \pi \omega_1^2 f(x)$. From the explicit form of $\chi''(\omega)$ it follows that the absorption of the energy of modulation field $H_m, \omega$ is of resonance character and is observed at the frequency $\omega = \Delta_0$ (if $\omega_0 > \Omega$) and of the frequency $\omega = -\Delta_0$ (if $\omega_0 < \Omega$); besides, at $\omega = \Delta_0$ frequency there occurs the energy absorption $(\chi''(\Delta_0) > 0)$, and at $\omega = -\Delta_0$ frequency – the energy radiation $(\chi''(-\Delta_0) < 0)$.

**Authors' address**:

**Michael D. Zviadadze**, E.Andronikashvili Institute of Physics, Tamarashvili 6, 0177 Tbilisi, Georgia
E-mail: m.zviadadze@mail.ru